\def\kb{\overline{k}}
\def\ob{\overline{\omega}}
\documentstyle[aps,prb,epsf,preprint,tighten]{revtex}
\begin{document}
\widetext
\title{Quantum phase transitions in spin systems\\
and\\
the high temperature limit of continuum quantum field theories}
\author{Subir Sachdev}
\address{Department of Physics, Yale University, \\
P.O. Box 208120, New Haven CT 06520-8120, U.S.A.}
\date{August 9, 1995}
\maketitle
\begin{center}
\tt Proceedings of the \\
19th IUPAP International Conference on Statistical Physics\\
Xiamen, China, July 31 - August 4 1995\\
World Scientific, to be published, edited by B.-L. Hao.
\end{center}

\begin{abstract}
We study the finite temperature crossovers in the vicinity of a zero
temperature quantum
phase transition. The universal crossover functions are observables of a
continuum
quantum field theory. Particular attention is focussed on the high temperature
limit
of the continuum field theory, the so-called ``quantum-critical'' region.
Basic features of crossovers are illustrated by a simple solvable model of
dilute
spinless fermions, and a partially solvable model of dilute bosons. The low
frequency
relaxational behavior of the quantum-critical region is displayed in the
solution of the
transverse-field Ising model. The insights from these simple models lead to a
fairly
complete understanding of the system of primary interest: the two-dimensional
quantum
rotor model, whose phase transition is expected to be
in the same universality class as those in antiferromagnetic
Heisenberg spin models. Recent work on the experimental implications of these
results for
the cuprate compounds is briefly reviewed.
\end{abstract}
\pacs{PACS: xxxxx}
\narrowtext

\section{Introduction}

Consider a quantum system on an infinite lattice described by the Hamiltonian
${\cal H}(g)$, with $g$ a dimensionless coupling constant. For any reasonable
$g$,  all observable properties of the {\em ground state\/} of ${\cal H}$
 will vary smoothly as $g$
is varied. However, there may be special points, like $g=g_c$, where there is
a non-analyticity in some property of the ground state: we identify $g_c$
as the position of a quantum phase transition. In finite lattices,
non-analyticities can only occur at level crossings; the possibilities in
infinite systems are richer as avoided level crossings can become sharp in the
thermodynamic limit. In this paper, I will restrict my discussion to second
order quantum transitions, or transitions in which the correlation length
and correlation time diverge as $g$ approaches $g_c$. As I will review
below, any such quantum transition can be used to define a continuum quantum
field theory (CQFT): the CQFT has no intrinsic short-distance (or
ultraviolet) cutoff.
% and has excitations with a spectrum $\omega (k)$ defined
% for all values of the momentum $k$.
The main purpose of this paper is to
review some recent work~\cite{jinwu,cs,csy,sss,conserve} on the properties
of ${\cal H} (g)$ at
{\em finite temperatures\/} ($T$) in the
vicinity of $g=g_c$. This is equivalent to a study of the finite $T$
crossovers of the associated CQFT. We shall focus especially on the
dynamic properties of a ubiquitous finite $T$ region, usually called
``quantum critical''~\cite{chn} (as we shall see below, there are reasons why
this
name is misleading and not quite appropriate; nevertheless, I will use it
here).
The quantum-critical region appears as the high $T$ limit of the CQFT; unlike
the
statics of classical lattice models, the high $T$ limit of a CQFT is usually
highly non-trivial.  All of this discussion will take place in the context of
some simple examples drawn from quantum spin systems.

I set the stage by reviewing the Wilsonian approach to critical phenomena and
field theories~\cite{brezin},
using the perspective of quantum critical phenomena. By the usual Trotter
product
decomposition, we can set up the partition function of ${\cal H}(g)$ as a
functional integral over degrees of freedom which fluctuate as a function of
the
spatial co-ordinate $x$ and imaginary time $\tau$.  Let us now examine the
behavior
of this functional integral under the rescaling
transformation~\cite{hertz,boyanovsky}
\begin{equation}
x \rightarrow e^{-\ell} x ~~~~~~~~~~~\tau \rightarrow e^{-z \ell} \tau
\label{rg}
\end{equation}
The dynamic exponent $z$ determines the relative scaling dimensions of space
and
time  co-ordinates. The critical point at $g=g_c$ is a fixed point
of (\ref{rg}), and $g-g_c$ is a relevant perturbation away from this point.
We have therefore the flow equation
\begin{equation}
\frac{dg}{d\ell} = \frac{1}{\nu} ( g - g_c)
\end{equation}
which defines the critical exponent $\nu$. (For simplicity I do not discuss the
case of fixed points with more than one relevant perturbation, as they can
be treated in a similar manner). In the long-distance, long-time limit, this
deviation from the critical point will be characterized by some {\em
renormalized\/}
energy scale, $G$. I emphasize that $G$ is a dimensionful parameter, expressed
in the laboratory units of energy, and directly measurable in an
experiment; a typical example would be an energy gap. In the vicinity of the
critical point, the renormalized energy scale $G$ will be related to the bare
coupling
$g$ by
\begin{equation}
G \sim \Lambda | g - g_c |^{z\nu}
\label{Gg}
\end{equation}
where $\Lambda$ is an ultraviolet cutoff, measured for convenience in the units
of
energy too. From the perspective of a field theorist,
 the CQFT associated with the quantum
critical point is now defined by taking the limit $\Lambda \rightarrow \infty$
at fixed $G$; from
(\ref{Gg}) we see that, because $z\nu > 0$, it is possible to take this limit
by tuning the bare
coupling $g$ closer and closer to the critical point as $\Lambda$ increases.
(A condensed matter physicist would take the complementary, but equivalent,
perspective
of keeping $\Lambda$ fixed but moving closer to criticality by lowering his
probe frequency
$\omega \sim G$).
The resulting CQFT then contains only the energy scale $G$. At finite
temperatures, there is a second energy scale  $T$ (using units in which $k_B =
1$);
its thermodynamic properties will then be a universal function of the only
dimensionless ratio available---$G/T$. It is the purpose of this paper to
review
recent work on the crossovers as a function of $G/T$ in a number of systems,
and to highlight the unusual properties of the heretofore unexplored
high-temperature, ``quantum critical'', limit of the CQFT, $T \gg G$.

It is now easy to see why the high $T$ limit of the CQFT can be non-trivial.
A conventional high $T$ expansions of the lattice model ${\cal H}$ proceeds
with the series
\begin{equation}
\mbox{Tr} e^{- {\cal H} / T} = \mbox{Tr} 1 -
\frac{1}{T} \mbox{Tr} {\cal H} + \frac{1}{2T^2} \mbox{Tr} {\cal H}^2+ \ldots
\label{highT}
\end{equation}
The successive terms in this series are well-defined and finite because of the
ultraviolet cutoffs provided by the lattice. Further, the series is
well-behaved provided $T$ is larger than all other energy scales; in particular
we
need $T \gg \Lambda$. In contrast, the CQFT was defined by the limit
$\Lambda \rightarrow \infty$ at fixed $T$, $G$, so the high $T$ limit of the
CQFT corresponds to the intermediate temperature range $G \ll T \ll \Lambda$
of the lattice model. It is not possible to access this temperature range
by an expansion as simple as (\ref{highT}), and more sophisticated techniques,
to be discussed here, are necessary. (One could also, of course, determine a
large number of terms in (\ref{highT}) and then use some Pad\'{e} extrapolation
methods to access the $T \ll \Lambda$ region: this method has been used by
Sokol {\em et. al.\/}~\cite{rajiv} and I will not discuss it here).

In contrast to the static properties, the dynamic properties  of ${\cal H}$ are
already non-trivial in the high $T \gg \Lambda$ limit of the lattice model.
Although one expects some sort of incoherent, dissipative dynamics, the damping
co-efficients cannot be determined directly---all the approaches used so far
are
essentially variants of the methods discussed by Moriya~\cite{moriya} and
Forster~\cite{forster}, and use a short-time expansion, coupled with an
ansatz for the spectral function, to extrapolate to the long-time limit. In
this
paper, we will discuss  the dynamics in the high $T$ limit of the CQFT, $G \ll
T \ll
\Lambda$, or the ``quantum-critical dynamics''.  The dynamics continues to be
dissipative and relaxational, but is not amenable to a description either by a
classical
Boltzmann equation for a dilute gas of quasiparticle excitations, or by a
classical
Langevin-like models of the types discussed in the classic review of Hohenberg
and
Halperin~\cite{halphoh}. However, as we shall show, the scaling structure of
the CQFT does
permit some progress to be made; indeed we will discuss below the complete
solution of the
quantum-critical dynamics of a simple spin model, including the exact
determination of a
damping co-efficient.

We will begin our discussion with a simple solvable model of spinless fermions
in
Section~\ref{spinferm}: this will allow introduction of the main concepts in a
very simple
setting. Section~\ref{sec:bose} will extend these results to a related but more
complex model
of dilute bosons in spatial dimensions $d<2$. An explicit solution of the
relaxational
dynamics of the quantum-critical region will be obtained in the discussion in
Section~\ref{isingsec} on the Ising model in a transverse field. The expository
analyses of
these toy models will lead in Section~\ref{secrotor} to the main system of
interest---the
$O(3)$ quantum rotor model in $d=2$. We will also review applications of these
results to
numerical simulations and experiments. Finally Section~\ref{concsec} will
conclude by
highlighting the main results and noting recent work on related subjects.

\section{Dilute gas of spinless fermions}
\label{spinferm}
Much of the physics I wish to discuss is displayed in a surprisingly simple
model of a dilute gas of spinless fermions at finite temperature: its scaling
forms
have a structure identical to those of much more complicated models.
The main shortcoming of the model is that the associated CQFT has no
interactions,
and there is therefore no relaxational behavior in the scaling functions.

Consider the following Hamiltonian:
\begin{equation}
{\cal H}_F = - t \sum_{<ij>} \left( c_i^{\dagger} c_{j} +
c_{j}^{\dagger} c_i - c_i^{\dagger} c_i - c_j^{\dagger} c_j \right) - \mu
\sum_i
c_i^{\dagger} c_i
\label{fermiham}
\end{equation}
where $c_i$ is a spinless fermion annihilation operator at the site $i$ of a
$d$-dimensional hypercubic lattice, and $<ij>$ are nearest neighbors.
There are no interactions in ${\cal
H}_F$, so it is trivially solvable. Consider the ground state of ${\cal H}_F$
as a
function of the dimensionless coupling constant $g=\mu /t$. For $g < g_c = 0$,
the ground state has no particles. There is a non-analytic onset in the
density of particles at $g=g_c$, signaling a quantum phase
transition. The bandwidth $\sim t$ of the fermions plays the role of the upper
cutoff in energy ($\Lambda$) for this transition, and the critical region
defining the applicability of a CQFT is roughly $T, \mu \ll t$. In fact, it is
not difficult
to determine the exact effective action of the CQFT (in units with $\hbar=1$):
\begin{equation}
{\cal L}_F = \int d^d x \int_0^{1/T} d\tau
\Psi_F^{\dagger} (x, \tau) \left( \frac{\partial}{\partial\tau} -
\frac{\nabla^2}{2m}
- \mu \right) \Psi_F (x, \tau)
\label{cqft1}
\end{equation}
where $\Psi_F (x_i ) = a^{-d} c_i $, $m \sim 1/(ta^2)$, and $a$ is the lattice
spacing. For this action we can identify by the usual methods the exponents
$\nu=1/2$ and $z=2$ at the $\mu=0$ critical point; taking the density $n =
\langle
\Psi_F^{\dagger}
\Psi_F \rangle$ as the order parameter, we get the exponents $\beta = d/2$ and
$\eta = d$.
Further, it is also
easy to see that all interactions are irrelevant at this critical point in all
dimensions
$d > 0$ (the least irrelevant interaction term, $|\Psi_F \nabla \Psi_F |^2$,
becomes relevant only for $d<0$). Finally, for the renormalized energy scale
measuring the deviation from the critical point, $G$, we may take $G=\mu$,
the bare chemical potential in ${\cal H}_F$. Note that there is no
non-universal
scale-factor in the relationship between $G$ and $\mu$---this is a consequence
of
the triviality of the critical exponents. More typical models with anomalous
exponents
will have non-universal scale-factors.

\begin{figure}
\epsfxsize=5in
\centerline{\epsffile{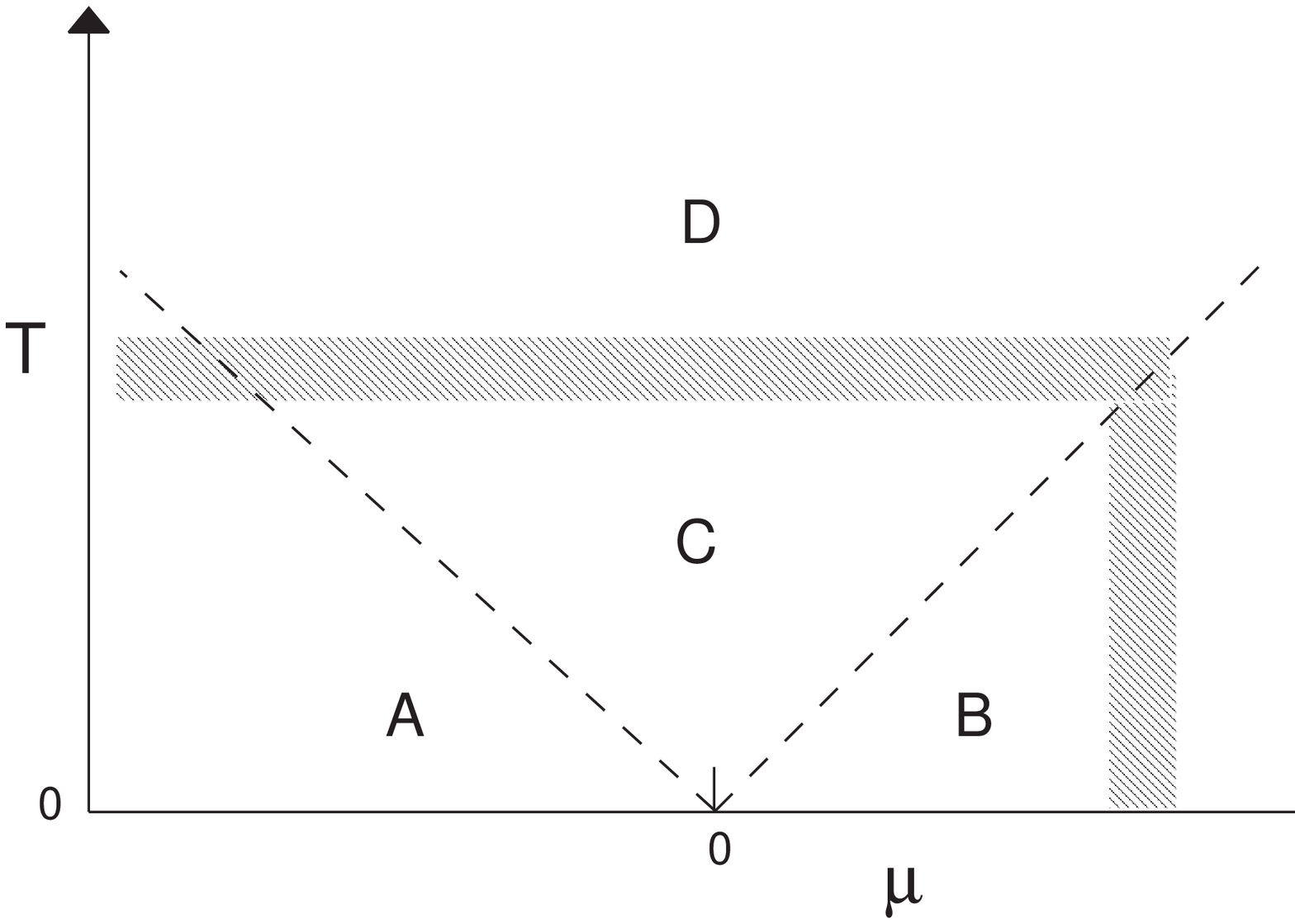}}
\vspace{0.1in}
\caption{Phase diagram of the dilute Fermi gas ${\cal H}_F$ (Eqn
(\protect\ref{fermiham}))
as a function of the chemical potential $\mu$ and the temperature $T$.
The regions A, B, C, are separated by crossovers; all observables in these
regions are
described  universal crossover functions of the CQFT ${\cal L}_F$ (Eqn
(\protect\ref{cqft1})).
Region A  has an exponentially activated fermion density, region B is a Fermi
liquid
(in $d=1$, a Luttinger liquid), and region C is quantum critical. The hatched
region marks the boundary of applicability of the CQFT and occurs
at $\mu, T \sim \Lambda = t$. The same phase diagram also applies to the dilute
Bose gas of Sec.~\protect\ref{sec:bose}, but the interpretation of the regions
is
different.}
\label{fermifig}
\end{figure}

We show in Fig~\ref{fermifig} the phase diagram of ${\cal H}_F$ as a function
of
$G = \mu$ and $T$.
Our interest is primarily in the regions within the hatched
lines where the CQFT (\ref{cqft1}) applies. Within this region there are three
physically distinct types of behavior (A, B and C): as ${\cal L}_F$ is
trivially solvable,
the universal properties of A, B, and C and the crossovers between them can be
exactly determined. Let us describe the regions in turn:

\noindent (A) {\em Activated} $\mu \ll -T$: The fermions are very dilute, with
a density
$\sim e^{\mu / T}$. Quantum effects are suppressed and the particles behave
classically.

\noindent (B) {\em Fermi or Luttinger liquid} $ \mu \gg T$: Now quantum
degeneracy effects are
paramount. At $T=0$, the ground state is a Fermi liquid (in $d=1$, a Luttinger
liquid);
at finite $T$
thermal effects lead to  a small number of particle and hole excitations near
the Fermi surface.

\noindent (C) {\em Quantum Critical} $|\mu| \ll T$: Unlike A and B, the
temperature
$T$ is the most important energy scale in this region. We can set $\mu = 0$
here
without much damage (all corrections involve positive powers of $\mu / T$). The
energy
of a typical excitation in this region is of order $T$ and as a result, quantum
and
thermal fluctuations are equally important.

The relationships between the regions becomes clearer upon considering an
explicit
example of a crossover function. The density of particles $n = \langle
\Psi_F^{\dagger}
\Psi_F \rangle$ obeys the scaling form
\begin{equation}
n = (2 m T)^{d/2} \Phi_n \left( \frac{\mu}{T} \right)
\label{phin}
\end{equation}
where the universal scaling function $\Phi_n (\overline{\mu})$ is given by
\begin{equation}
\Phi_n (\overline{\mu}) = \int \frac{d^d k}{(2 \pi)^d} \frac{1}{e^{k^2 -
\overline{\mu}} + 1}.
\end{equation}
Notice that there are no arbitrary scale-factors in (\ref{phin}).
In the activated region A ($\overline{\mu} \ll -1$), we have $\Phi_n \approx
e^{\overline{\mu}}
/(4 \pi)^{d/2}$, which
is  exactly the result we would have obtain from the classical
Maxwell-Boltzmann
statistics. In the Fermi liquid region B ($\overline{\mu} \gg 1$), $\Phi_n
\approx
\overline{\mu}^{d/2} V_d / (2 \pi)^d$, where $V_d$ is the volume of the unit
sphere in $d$
dimensions; this is the fully quantum result obtained by filling up the Fermi
sphere.
Most interesting is the quantum critical region C ($|\overline{\mu}| \ll 1)$
where
\begin{equation}
\Phi_n (\overline{\mu}) = \zeta \left( \frac{d}{2} \right) \left( \frac{\pi}{2}
\right)^{d}
\left( 1- 2^{1-d/2} \right) + {\cal O} (\overline{\mu})
\end{equation}
The value of $\Phi_n (0)$ depends upon the details of the Fermi distribution
function,
and not just its forms in the classical and quantum limits: this illustrates
our
assertion that quantum and classical effects are equally important in region C.

It is also interesting to compare the behavior of the density in the universal
region
C with the true high temperature limit
of the lattice model - region D. In other words, we are going vertically
upwards
in $T$ at $\mu = 0$ in Fig~\ref{fermifig}. It is easy to compute:
\begin{equation}
n = \left\{
\begin{array}{cc}
(2 m T)^{d/2} \Phi_n (0)  & |G| \ll T \ll \Lambda~~(\mbox{region C})\\
1/2a^d - c_1 / T  & T \gg \Lambda~~(\mbox{region D})
\end{array}
 \right.
\end{equation}
where $c_1$ is a non-universal constant (recall that in this model $G=\mu$
and $\Lambda = t$). Notice the difference between the universal high $T$ limit
of the CQFT (the first result) and the lattice high $T$ limit. I hope that this
example has illustrated the general principle, and in the remainder of the
paper
I will make no further reference to non-universal regions like D. It will be
implicitly assumed that we are working with the universal continuum theory,
and we will describe only regions like A, B and C.

Before closing our discussion on this deceptively simple model, we discuss the
scaling
form of observables as a function of momentum $k$ and frequency $\omega$.
Because of the absence of interactions, the single-particle Green's function is
trivial;
so we discuss the density-density correlator $\chi_n (k, \omega )$ which has a
slightly
more interesting structure. As the particles are free, $\chi_n $ is of course
given simply
by the Lindhard function, which can be manipulated into the scaling form
\begin{equation}
\chi_n ( k , \omega ) = \frac{(2m T)^{d/2}}{T} \Phi_{\chi_n} \left (
\frac{k}{\sqrt{2 m T}} , \frac{\omega }{T}, \frac{\mu}{T} \right)
\label{phichin}
\end{equation}
where $\Phi_{\chi_n}$ is a universal complex-valued function related to the
Lindhard
function. Notice that, like the scaling form (\ref{phin}), there are again no
arbitrary
scale factors. Further, $\Phi_{\chi_n}$ is well-defined at $\mu/T = 0$, where
it
yields the dynamic susceptibility of the quantum-critical region C. We will see
several
other examples of scaling forms like (\ref{phin}) and (\ref{phichin}) in this
paper,
but the scaling functions will not be as simple as they are here.

\section{Dilute Bose gas}
\label{sec:bose}
Now we consider the same density onset quantum transition considered in
Sec~\ref{spinferm},
but for the case of bosons. The discussion here is drawn from that of Sachdev,
Senthil and
Shankar~\cite{sss} to which the reader is referred for further details.

Unlike the case for spinless fermions, it is no longer possible to ignore the
interactions
between the particles. We consider the properties of the following continuum
model
\begin{eqnarray}
{\cal L}_B &=&
\int d^d x \int_0^{1/T} d\tau
\Psi_B^{\dagger} (x, \tau) \left( \frac{\partial}{\partial\tau} -
\frac{\nabla^2}{2m}
- \mu \right) \Psi_B (x, \tau) \nonumber \\
&+& \frac{1}{2} \int d^d x \int d^d x' \int_0^{1/T} d\tau
| \Psi_B ( x, \tau) |^2 v(x- x') |\Psi_B ( x', \tau) |^2
\label{cqft2}
\end{eqnarray}
where $\Psi_B$ is a boson annihilation operator, and $v(x)$ is a repulsive
interaction of
range $\sim a$. Like ${\cal L}_F$, ${\cal L}_B$ has a quantum phase transition
at $\mu=T=0$, and we will discuss its universal properties here. Because of the
finite
range of $v$, the universality only sets in at distances larger than $a$.
A straightforward RG analysis~\cite{fisher} of the vicinity of the quantum
critical point
shows that $v$ flows into the $v=0$ fixed point for $d\geq 2$. It turns out
that
$v$ is actually dangerously irrelevant for $d\geq 2$: we do not wish to enter
into a
discussion of such effects here, and so most of our remaining discussion will
be
restricted to $d < 2$.

For $d<2$, $v(x)$ flows into a universal fixed point
interaction $v(x) = u^{\ast} \delta(x)$
for some $d$ dependent constant $u^{\ast}$.
The scaling structure of this fixed point turns out to be
very closely related to the non-interacting spinless fermion model of
Sec~\ref{spinferm}
for the same value of $d$. All exponents and scaling forms of the boson and
fermion
models are identical, but the scaling functions themselves are different.
The crossover phase diagram of the $d<2$ dilute Bose gas is essentially
identical to
the fermion phase diagram in Fig~\ref{fermifig}, but the physical
interpretation of
the phases is somewhat different:

\noindent (A) {\em Activated} $\mu \ll -T$: This is essentially identical to
the
fermion case as the particles are dilute and their quantum statistics plays a
negligible role.

\noindent (B) {\em Incipient Superfluid} $ \mu \gg T$: The ground state
is now a  superfluid (in $d=1$ a Luttinger liquid),
but classical thermal fluctuations destroy the long range order at any non-zero
temperature. Nevertheless, the phase coherence length is large and system
behaves
like a superfluid at short scales.

\noindent (C) {\em Quantum Critical} $|\mu| \ll T$: This is similar to the
fermion case
in that $T$ is the most important energy scale. However, there are now strong
interactions
among the particles, leading to an incoherent excitation spectrum. The system
does not
display characteristics of a superfluid ground state at any length scale,
but instead crosses over directly from free particle behavior at short time
scales,
to dissipative, relaxational dynamics at long time scales.
The single particle Green's function $G(x, \tau) = \langle \Psi_B (x,\tau)
\Psi_B^{\dagger} (0,0) \rangle$ has a non-trivial scaling function $\Phi_G$
(this form
actually holds in all three regions A, B, and C)
\begin{equation}
G^R (k, \omega ) = \frac{1}{T} \Phi_G \left (
\frac{k}{\sqrt{2 m T}} , \frac{\omega }{T}, \frac{\mu}{T} \right).
\label{phiG}
\end{equation}
We have Fourier transformed and analytically continued to the retarded Green's
function
at real frequencies. This scaling form also holds for the spinless fermions of
Section~\ref{spinferm}, but the scaling function then is simply the free
fermion form $\Phi_G
(\kb,\ob,
\overline{\mu}) =  1/(\ob - \kb^2 - \overline{\mu} +
i\eta)$ where
$\eta$ is a positive infinitesimal. Computing
$\Phi_G$, and other scaling functions, for bosonic quantum critical dynamics is
not
as easy. One approach~\cite{sss} is to expand in powers of the fixed-point
interaction
$u^{\ast}$ which becomes small as $d$ approaches 2 from below: this becomes an
expansion in $\varepsilon=2-d$.

In $d=1$, it is possible to make more explicit progress. It has been
argued~\cite{sss} that now
$u^{\ast} = \infty$. The bosons thence become impenetrable, and their quantum
mechanics
becomes identical to those of free fermions. Hence, in $d=1$, the
quantum critical dynamics of dilute gases of spinless fermions and boson are
described
by the same CQFT, ${\cal L}_F$ of Eqn (\ref{cqft1}). For the bosonic system we
have to supplement
${\cal L}_F$ with the following non-local relationship between the boson and
fermion operators
(essentially a continuum Jordan-Wigner transformation)
\begin{equation}
\Psi_B (x ) = \exp \left( i \pi \int_{-\infty}^{x} dx'
\Psi_F^\dagger (x') \Psi_F (x') \right) \Psi_F (x)
\label{jw}
\end{equation}
We are not home yet, as
evaluating correlators of (\ref{jw}) under (\ref{cqft1}) is not easy. Korepin
and
Slavnov~\cite{korep,korepbook} have succeeded in showing how this problem may
be reduced to
determining the solution and Fredholm determinant of a linear Fredholm integral
equation. At
this stage, numerical analysis is required, and some scaling functions have
been determined
to essentially arbitrary accuracy~\cite{sss}.

Finally, an additional comment about the $d \geq 2$ case.
There is now true superfluidity and a finite temperature phase transition to a
normal state,
all within region B. The remainder of the phase diagram remains the same as in
the $d\leq 2$
case.

\section{Ising Model in a transverse field}
\label{isingsec}
Unlike the dilute Fermi and Bose gases, the Ising model
possesses anomalous exponents. Yet it is simple enough in $d=1$ to allow exact
computation
of a quantum-critical dynamic correlation function. We will also review, in
this section, the
very useful and general mapping between the quantum model and an equivalent
classical statistical
mechanics model; we will then discuss the crossovers in the phase diagram like
Fig~\ref{fermifig} in the context of the classical model. Some of
of the following discussion is
a review of well-known properties of the Ising
model~\cite{suzuki,kogut,drouffe,ising}; our main
purpose here is to use the explicit solution in $d=1$ to present a physical
interpretation which
generalizes to other quantum phase transitions.

We will explicitly discuss the following Hamiltonian, describing the Ising
model in a
transverse field in $d=1$ (we will remark briefly on the generalization to
higher $d$):
\begin{equation}
{\cal H}_I = - J \sum_i \left(  g \sigma_{x,i} + \sigma_{z,i} \sigma_{z,i+1}
\right)
\label{hamising}
\end{equation}
where $J > 0$ in an overall energy scale, $g > 0$ is a dimensionless coupling
constant,
and $\sigma_{x,i}, \sigma_{z,i}$ are Pauli matrices on a chain of sites, $i$.
Consider the ground states of ${\cal H}_I$ for small and large $g$ in turn. For
small
$g$, the second term in (\ref{hamising}) dominates and the spins all align
themselves either
in the $+z$ or $-z$ directions: there is a spontaneous magnetization and
spin-reversal
symmetry is broken. On the other hand, for large $g$, the first term in
(\ref{hamising})
prefers a state which is in a superposition of $\sigma_z$ eigenstates, with
different sites
uncorrelated: the wavefunction looks like $\prod_i ( |+\rangle_i + |-\rangle_i
)$.
These two limits are separated by a phase transition at $g = g_c$ (in fact,
$g_c = 1$,
exactly, because of a self-duality property of (\ref{hamising})). In this
section, we shall
discuss the finite $T$ dynamic properties in the vicinity of $g=g_c$.

\begin{figure}
\epsfxsize=5in
\centerline{\epsffile{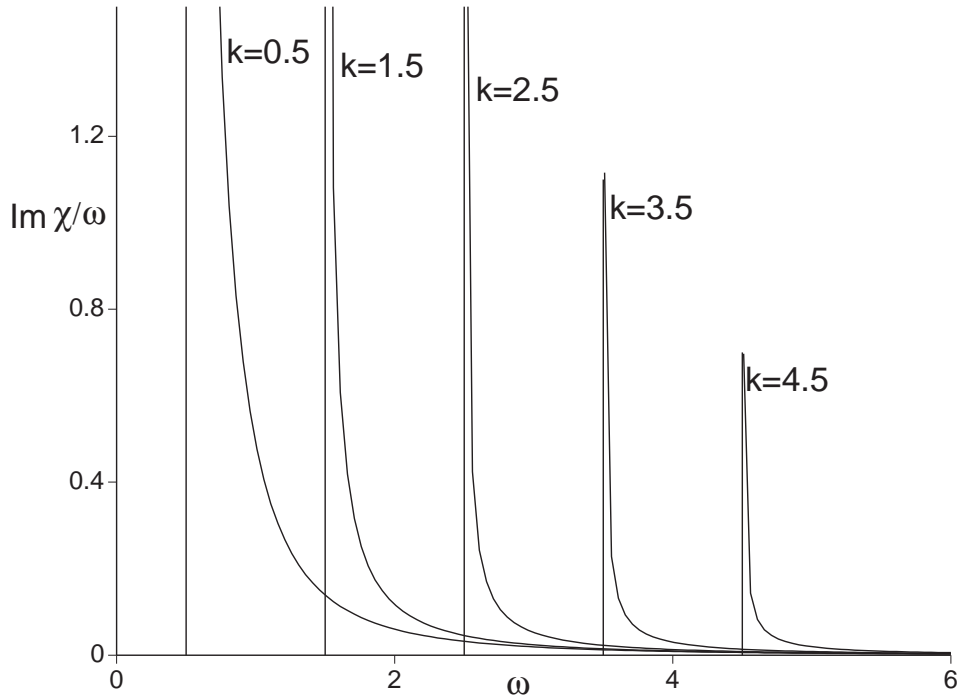}}
\vspace{0.1in}
\caption{Spectral density of the transverse field Ising model
(\protect\ref{hamising})
at its critical point $g=g_c$ at $T=0$. We chose $Z=c=1$ in Eqn
(\protect\ref{chiteq0})}
\label{chiteq0fig}
\end{figure}

To begin, we recall the well-known fact~\cite{suzuki,kogut} that the $T=0$
correlators of ${\cal H}_I$
are similar to those in the classical, two-dimensional, Ising model given by
the partition
function
$\mbox{Tr}~e^{-F}$ with
\begin{equation}
F = - K \sum_{<ij>} \sigma_{z,i} \sigma_{z,j}
\end{equation}
where the sites $i,j$ now lie on a square lattice (say).
This classical model has a phase transition at $K=K_c$.
At the level of the critical
continuum theories, the mapping between the classical and quantum models become
exact~\cite{kogut}.
There is a simple relationship between the  two-dimensional field theory, with
classical
degrees of freedom, describing
$F$ in the vicinity of
$K=K_c$ and the one-dimensional quantum field theory describing ${\cal H}_I$
near
$g=g_c$: one simply identifies one of the spatial directions of the classical
theory as an
imaginary time, and then analytically continues the correlators to real time,
to obtain
observables of the quantum theory. This mapping leads immediately to some
useful information.
As the classical theory is spatially isotropic, the quantum theory has dynamic
exponent
$z=1$. Further, in the classical model the $\sigma_z~\sigma_z$ correlator
behaves like
$\sim p^{-7/4}$ in momentum space~\cite{drouffe} ($p$ is the two-dimensional
momentum of the
classical model); analytically continuing this to real time, we obtain for the
dynamic
susceptibility
$\chi (k,
\omega )$ (this is the
$\sigma_z~\sigma_z$ correlator of the quantum ${\cal H}_I$ and $k$ is now a
one-dimensional
spatial momentum) as a function of
$d=1$ momentum
$k$ and frequency
$\omega$:
\begin{equation}
\chi (k, \omega ) = \frac{Z}{(c^2 k^2 - \omega^2)^{7/8}}~~~~~~T=0, g=g_c.
\label{chiteq0}
\end{equation}
Here $Z$ and $c$ (an excitation velocity) are non-universal constants. We plot
$\mbox{Im}
\chi (k,\omega )/\omega$ in  Fig~\ref{chiteq0fig}.
Notice that there are no delta functions in the
spectral density, indicating the absence of any well-defined quasiparticles.
Instead, we have
a critical continuum of excitations. We can also compute the dynamic local
susceptibility,
\begin{equation}
\chi_L^{\prime\prime} (\omega ) = \int \frac{dk}{2\pi} \mbox{Im} \chi (k,
\omega),
\end{equation}
a quantity often measured in neutron scattering experiments:
\begin{equation}
\chi_L^{\prime\prime} ( \omega ) = \mbox{sgn} (\omega) \frac{\sqrt{\pi} Z}{2
\Gamma(7/8)
\Gamma(5/8) c}
\left(\frac{1}{|\omega|}\right)^{3/4}~~~~~~T=0, g=g_c.
\label{chilteq0}
\end{equation}
This quantity is the density of states of local spin-flip excitations, and has
a divergence as $\omega \rightarrow 0$.

Our discussion so far has been at $T=0$, and let us turn now to non-zero $T$.
A finite $T$ translates into a {\em finite size} $L_\tau = 1/T$
for the classical model along imaginary time direction. Periodic boundary
conditions are
imposed along the finite direction, and so the classical model has the geometry
of
a cylinder with circumference $L_\tau = 1/T$. We now discuss the crossovers at
finite $T$
in the context of the classical model. At $L_\tau = \infty$ ($T=0$) we can
characterize
the deviations from criticality by the correlation length $\xi \sim
|g-g_c|^{-1}$ (for the
quantum model we may choose our renormalized energy scale $G=1/\xi \sim
|g-g_c|$).
At distances shorter than $\xi$, the spins display  critical correlations
characteristic
of the point $K=K_c$;
it is only at distances larger than $\xi$ that they become sensitive to the
value
of $K-K_c$ and display characteristics of the ordered (for $K > K_c$) or the
paramagnetic (for $K < K_c$) phase. Now consider the effect of a finite
$L_\tau$; there are
two distinct possibilities:

\noindent ({\em i}) $\xi < L_\tau$ (for the quantum model, $T < G$): Moving
from the shortest
to  largest length scales, the crossover from critical to non-critical (either
ordered
or paramagnetic) behavior still occurs at a scale $\sim \xi$; the
length scale $L_\tau$ has little effect at this point. The effects of $L_\tau$
only become apparent
at larger scales, at which point it is permissible to use an effective model
which
characterizes the non-critical ground state.

\noindent ({\em ii}) $L_\tau < \xi$ (for the quantum model, $T > G$):
Now the short-distance critical fluctuations see a finite size $L_\tau$ {\em
before}
they have had a chance to become sensitive to $K-K_c$. These critical
fluctuations
are quenched by finite size effects in a universal way. The resulting
non-critical
theory then responds only weakly at the scale $\xi$. Note that the system does
not
display characteristics of the ordered or the paramagnetic state, of the
$L_\tau=\infty$
system, at {\em any\/} length scale.

The above arguments are summarized in Figs~\ref{isingfig}
and~\ref{isingomegafig}; notice that
Fig~\ref{isingfig} is quite similar to Fig~\ref{fermifig}.
In all three regions of
Fig~\ref{isingfig}, at the largest frequencies, $\omega$, ${\cal H}_I$ displays
the critical
correlations of the
$g=g_c$ point as depicted in Fig~\ref{isingomegafig} (we have returned now to
the language of the
quantum model ${\cal H}_I$).

In regions A and B there is a crossover from these critical fluctuations, at
an energy scale $G$, to the behavior of the ordered ($g<g_c$) or paramagnetic
($g>g_c$)
ground state of ${\cal H}_I$ (see Fig~\ref{isingomegafig}). Both ground states
have a gap, and thermal
fluctuations will lead to dilute gas of quasiparticle excitations. We expect
that an effective
classical model (like Glauber~\cite{glauber} dynamics, which is similar in
spirit to the
Langevin models of Hohenberg and Halperin~\cite{halphoh}) will provide a
suitable description of these thermal fluctuations. In $d=1$, on the ordered
side ($g < g_c$), these
quasiparticle excitations are the `kink' and `anti-kink' solitons; even an
infinitesimal concentration
of these is sufficient to destroy long-range order at any finite temperature.
However,
for $d>1$, long-range order is not immediately destroyed: as a result there is
finite
temperature phase transition within region A where the magnetic moment
disappears;
this transition will be in the universality class of the $d$-dimensional
classical Ising
model~\cite{suzuki}.

In the quantum-critical region C, the critical fluctuations are quenched by
thermal
effects at the energy scale $T$ (see Fig~\ref{isingomegafig}; an early analysis
of finite $T$
crossovers in the transverse-field Ising model by Suzuki~\cite{suzuki} failed
to identify region C).
This quenching is completely universal and will be described explicitly below.
The system has had no
chance to display any characteristic of either non-critical ground state at any
frequency scale.
At low frequencies, the system realizes a new quantum relaxational regime. It
is this
regime which is really characteristic of the region C, and not the high
frequency
critical behavior which is present in all three regions. It is in this sense
that
the name ``quantum-critical'' of region C is a misnomer.

As in the case of the dilute Fermi and Bose gases considered earlier,
the dynamic susceptibility $\chi(k,
\omega)$ will satisfy a universal scaling form over the regions of
Fig~\ref{isingfig}:
\begin{equation}
\chi(k, \omega) = \frac{Z}{T^{7/4}} \Phi_\chi \left( \frac{ck}{T},
\frac{\omega}{T},
\frac{G}{T} \right)
\label{chiscale}
\end{equation}
In the following we will determine the leading term in $\Phi_\chi$ in the
quantum-critical region C {\em i.e.} we will present an exact expression for
$\Phi_\chi ( \kb, \ob, 0)$.

In the classical model $F$ at $K=K_c$ and $L_\tau = \infty$,
we know from (\ref{chiteq0}) that
the Green's function
$G(x, \tau) = \langle \sigma_z (0,0) \sigma_z (x, \tau) \rangle \sim (x^2 +
\tau^2
c^2)^{-1/8}$; we are using the label $\tau$ for spatial direction corresponding
to
imaginary time. We can now use a remarkable result of Cardy~\cite{cardy}, which
relies on the
conformal invariance of this critical theory, to obtain an exact result for $G$
in a system with a finite $L_\tau$:
\begin{equation}
G(x, \tau) = \frac{\Gamma (1/8) Z}{2^{13/8} \pi^{3/4} \Gamma (7/8) c}
\left(\frac{1}{L_{\tau}}\right)^{1/4} \left(\frac{1}{\cosh(2\pi x/L_\tau c) -
\cos(2\pi\tau/L_\tau)}\right)^{1/8}
\label{Gxt}
\end{equation}
This result has been asserted earlier by an inspection of the partial
differential
equation satisfied by $G$~\cite{luther}. Although results like (\ref{Gxt}) have
been known to conformal
field theorists for some time, they usually interchange the roles of $x$ and
$\tau$ {\em i.e.} they
consider systems of finite spatial length $L_x$, and infinite temporal length,
so that the system
is in its ground state. For us, the spatial extent is infinite, and there is a
finite
length $L_\tau$ along the $\tau$ direction, with the correlator (\ref{Gxt})
 periodic in
$\tau$ with period
$L_\tau$ (such a perspective has also been discussed by Shankar~\cite{shankar}
and
by Korepin {\em et. al.}~\cite{korepbook}).
Note that it doesn't really make sense to talk about the long imaginary time
limit,
$\tau \gg L_\tau$. However, after analytic continuation to real time, the long
time limit
of the quantum problem, $t \gg 1/T$ or $\omega \ll T$, is eminently sensible,
and is
precisely the new quantum relaxational regime that we wish to access.
The analytic continuation is a little more convenient in Fourier space:
we Fourier transform (\ref{Gxt}) to obtain $G(k, \omega_n)$ at the Matsubara
frequencies $\omega_n$ and then analytically continue to real frequencies
(there are some
interesting subtleties in the Fourier transform to $G(k, \omega_n)$ and its
analytic structure
in the complex $\omega$ plane, which are discussed elsewhere~\cite{sss}). This
gives
us the universal function $\Phi_\chi$ in the
quantum-critical
region:
\begin{equation}
\Phi_\chi (\kb , \ob, 0) = \frac{1}{(4 \pi )^{7/4}}
\frac{\displaystyle \Gamma \left( \frac{1}{16} + i \frac{\ob + \kb}{4 \pi}
\right)
\Gamma \left( \frac{1}{16} + i \frac{\ob - \kb}{4 \pi} \right)}
{\displaystyle \Gamma \left( \frac{15}{16} + i \frac{\ob + \kb}{4 \pi} \right)
\Gamma \left( \frac{15}{16} + i \frac{\ob - \kb}{4 \pi} \right)}.
\label{phichi}
\end{equation}
We show a plot of $\mbox{Im} \Phi_\chi /\ob$ in Fig~\ref{chifig}.
This result is
the finite $T$ version of Fig~\ref{chiteq0fig}. Notice that the sharp features
of Fig~\ref{chiteq0fig} have been smoothed out on the scale $T$, and there is
non-zero absorption at
all frequencies. We can also observe the crossover as a function of frequency
claimed
earlier in Fig~\ref{isingomegafig}. Notice that for $\ob, \kb \gg 1$ there is a
well-defined
peak in $\mbox{Im} \Phi_\chi /\ob$ (Fig~\ref{chifig}) rather like the $T=0$
critical behavior
of Fig~\ref{chiteq0fig}. However, for $\ob, \kb \ll 1$ we cross-over to the
quantum relaxational
regime and the spectral density $\mbox{Im} \Phi_\chi /\ob$ is similar to a
Lorentzian
around $\ob = 0$. This relaxational behavior can be characterized by a
relaxation
rate $\Gamma_R$ defined as~\cite{halphoh}
\begin{equation}
\Gamma_R^{-1} = -i \left.
\frac{\partial \ln \chi (0, \omega)}{\partial \omega} \right|_{\omega =
0};
\label{gammadef}
\end{equation}
(this is motivated by the phenomenological relaxational form $\chi(0, \omega) =
\chi_0 / (1 - i \omega / \Gamma_R + {\cal O}(\omega^2))$).
{}From (\ref{chiscale}) and (\ref{phichi}) we determine:
\begin{equation}
\Gamma_R = \left( 2 \tan \frac{\pi}{16} \right) \frac{k_B T}{\hbar},
\label{gammares}
\end{equation}
where we have returned to physical units. The ease with which this result was
obtained
belies (I claim) its remarkable nature. Notice that we are working in a closed
Hamiltonian
system, evolving unitarily in time with the operator $e^{-i{\cal H}_I t}$, from
an
initial density matrix given by the Gibbs ensemble at a temperature $T$. Yet,
we
have obtained relaxational behavior at low frequencies, and determined
an exact value for a dissipation constant. Such behavior is more typically
obtained in
phenomenological models which couple the system to an external heat bath and
postulate an equation of motion of the Langevin type. Notice also that ${\cal
H}_I$
in (\ref{hamising}) is known to be integrable with an infinite number of
conservation
laws~\cite{kogut}.
However, the conservation laws are associated with a mapping to a free fermion
model
and are highly non-local in our $\sigma_z$ degrees of freedom; they play
essentially no role
in our considerations, and do not preclude relaxational behavior in the
$\sigma_z$
variables.

For completeness we also present results on a related observable which shows
the crossover from critical to quantum relaxation behavior. We consider the
local
susceptiblity $\chi_L^{\prime\prime}$, and
obtain the finite $T$ form of (\ref{chilteq0}) by integrating (\ref{phichi})
over momenta:
\begin{eqnarray}
\chi_L^{\prime\prime} ( \omega ) &=&
\frac{Z}{c T^{3/4}} \Phi_L \left(  \frac{\omega}{T},
\frac{G}{T} \right) \label{phildef} \\
\Phi_L ( \ob, 0) &=& \frac{1}{2^{7/4} \pi^{5/4} \Gamma (5/8) \Gamma(7/8)}
\sinh \left( \frac{\ob}{2} \right) \left| \Gamma \left(
\frac{1}{8} - i \frac{\ob}{2 \pi} \right) \right|^2
\end{eqnarray}
A plot of the scaling function $\Phi_L$ is shown in Fig~\ref{chilocfig}.
The function has
the asymptotic limits:
\begin{equation}
\Phi_L (\ob, 0) = \left\{
\begin{array}{cc}
\displaystyle
\frac{\Gamma^2 (1/8)}{2^{11/4} \pi^{5/4} \Gamma (5/8) \Gamma(7/8)} \ob &
|\ob| \ll 1 \\
\displaystyle \mbox{sgn} (\ob) \frac{\sqrt{\pi}}{2 \Gamma(7/8)
\Gamma(5/8) }
\left(\frac{1}{|\ob|}\right)^{3/4} & |\ob| \gg 1
\end{array}
\right.
\end{equation}
The small $\ob$ behavior is relaxational as $\Phi_L$ is linear in frequency,
and the critical behavior at large frequencies agrees with (\ref{chilteq0}).

\section{Quantum rotors in two dimensions}
\label{secrotor}

The study of this model is of direct experimental interest, as it is
believed~\cite{chn} to be
a reasonable model of the spin fluctuations in antiferromagnetic compounds like
$La_2 Cu O_4$ and its lightly doped variants. The insight gained from the
simple models studied in
the previous sections will now be of great use, and we will rapidly be able to
present a
scaling analysis of its quantum phase transition.

The Hamiltonian of the quantum rotor model is
\begin{equation}
{\cal H}_R =  J \sum_i \frac{g}{2} \vec{L}_i^2 - J \sum_{<ij>} \vec{n}_i .
\vec{n}_j
\end{equation}
where $J > 0 $ is an overall energy scale, $g>0$ is a dimensionless coupling
constant,
and $i$,$j$ are the sites of a two dimensional lattice ($<ij>$ denotes nearest
neighbors).
Notice the similarity between the forms of ${\cal H}_R$ and ${\cal H}_I$ in
(\ref{hamising}): it
will turn out that the corresponding terms play a similar role. On each site
$i$ of the lattice
we have the 3-component vector operators $\vec{L}$, $\vec{n}$ (dropping the
site index),
which obey the commutation relations:
\begin{equation}
[n_a , n_b] = 0~~,~~[L_a , n_b] = i \epsilon_{abc} n_c~~,~~[L_a , L_b] = i
\epsilon_{abc} L_c .
\end{equation}
The vector $\vec{n}$ is of unit length $\vec{n}^2 = 1$, and its orientation
identifies
direction of the local magnetic order; the quantum rotor model is usually
considered as an
effective model for an underlying system of Heisenberg spins---in this case the
magnetic
order can be any ordering which is specified by a single vector and has no
spatially
averaged magnetic moment. The simplest example of this is the two sublattice
N\'{e}el ordering, and
we will therefore refer to $\langle \vec{n} \rangle$ as the N\'{e}el order
parameter. The $\vec{L}$
operator measures the angular momentum, and as all phases have no net magnetic
moment,
we will always have $\langle \vec{L} \rangle = 0$.

For further insight into the meaning of ${\cal H}_R$, consider the eigenstates
of a
single site Hamiltonian $J g \vec{L}^2 /2$. This describes a particle moving on
a unit
sphere with angular co-ordinate $\vec{n}$ and kinetic energy $Jg \vec{L}^2 /2$.
Its eigenenergies
are $J g \ell (\ell +1 )/2$ with degeneracy $2 \ell + 1$ where $\ell = 0, 1, 2,
3\ldots$.
The ground state is a non-degenerate singlet ($\ell = 0$) and has maximum
uncertainty in
the orientation of $\vec{n}$. For large $g$, the ground state of ${\cal H}_R$
can be approximated
by the tensor product of $\ell=0$ states on each site. This state is clearly a
quantum paramagnet
and has a gap, $\Delta$, to all excitations. Notice the similarity between this
state
and the large $g$ quantum paramagnet of the Ising model ${\cal H}_I$; in both
cases the order
parameters $\sigma_z$, $\vec{n}$ are in a state of maximum uncertainty. The
small $g$ limit
of ${\cal H}_R$ is also similar to the small $g$ limit of ${\cal H}_I$: now the
exchange
interactions between the sites prefer a state in which $\vec{n}$ has the same
definite orientation on
each site. Therefore, we expect long-range N\'{e}el order in the small $g$
ground state.
These two limiting states will be separated by a quantum phase transition at
$g=g_c$,
which is, of course, the main subject of interest in this section.

Before discussing the critical properties, we pause to remark on the
relationship between the rotor
model and Heisenberg antiferromagnets. Consider a pair of antiferromagnetically
coupled
spin-$S$ Heisenberg spins: the eigenstates of this pair will have energies
$\propto \ell (\ell + 1) - \mbox{const}$ for $\ell = 0, 1, \ldots 2S$. Notice
the similarity
between these states and those of a single quantum rotor; the only difference
is that there
is no upper limit on the maximum value of $\ell$ in the rotor case. However, it
is reasonable
to expect that these extra high energy states will not modify the low energy
properties of lattice
models. Therefore, there is little reason to doubt that the critical properties
of Heisenberg
antiferromagnets with a natural pairing of spins (or more generally, a natural
clustering
into an even number of spins) will be same as those of the rotor model.
Antiferromagnets with no such pairing contain net Berry phase terms in their
imaginary time path
integral, beyond those present for the rotor model. However these Berry phases
cancel between
the sites, except for ``hedgehog''-like spacetime singularities~\cite{berry},
and it has been
argued~\cite{jinwu,csy} that these remnant Berry phases have no effect of the
leading critical
singularities. The reader is referred to the original papers for further
discussion on these subtle
issues~\cite{rs,csy}: we will restrict our discussion here to the much simpler
rotor model ${\cal H}_R$.

The critical properties and phase diagram of the $d=2$ rotor model ${\cal H}_R$
turn out to
be remarkably similar to those of the transverse field Ising model ${\cal H}_I$
of (\ref{hamising})
in $d=1$. Like the $d=1$ Ising model, the $d=2$ rotor model has no phase
transition at
any finite $T$: so the phase diagram of Fig~\ref{isingfig} applies to ${\cal
H}_R$, with no
phase boundary in region A (the phase diagram for ${\cal H}_R$ in $d=2$ was
obtained
first by Chakravarty {\em et. al.}~\cite{chn}). In both
systems, the
$T=0$,
$g=g_c$ critical point has
$z=1$. In the case of
${\cal H}_I$ this critical point was described by a CQFT which upon analytic
continuation to
imaginary time was the field theory of the two-dimensional classical Ising
model. The analogous
mapping for ${\cal H}_R$ yields the CQFT associated with the field theory for
the
three-dimensional, classical, Heisenberg ferromagnet. The universal scaling
functions
describing the crossovers in Fig~\ref{isingfig} and~\ref{isingomegafig} have an
identical form in
both theories, although, because the critical field theories are different, the
critical exponents,
universal amplitude ratios, and  scaling functions will have different
numerical values.
The explicit results presented in
Sec~\ref{isingsec} for quantum-critical region C of the Ising model, all apply,
unchanged in form,
to the region C of the $d=2$ rotor model: the qualitative features of the
spectral functions
in Figs~\ref{chiteq0fig},~\ref{chifig}, and~\ref{chilocfig} remain the same,
and the relaxation
rate $\Gamma_R$ (defined in (\ref{gammadef})) satisfies (\ref{gammares}) but
with a different
universal numerical prefactor.  However, unlike the $d=1$ Ising model, we
cannot now get exact
numerical results for the scaling functions of ${\cal H}_R$: this is because
the three-dimensional
classical Hiesenberg ferromagnet is not exactly solvable (unlike the
two-dimensional classical
Ising model). Instead we have to be satisfied by approximate methods;
reasonably accurate
numerical estimates can be obtained in the $1/N$ expansion which has been
discussed at length
by Chubukov, Sachdev and Ye~\cite{csy}.

There is a small, but significant, difference between the Ising model in $d=1$
and the rotor
model in $d=2$ which cannot go unmentioned. This difference applies mainly to
``ordered'' regime in region A (See Figs~\ref{isingfig}
and~\ref{isingomegafig}). The $T=0$
ground state of ${\cal H}_I$ for $g<g_c$ has a gap, associate with the finite
energy cost
of creating a kink or anti-kink soliton. In contrast, the $g < g_c$ ground
state
of ${\cal H}_R$ has gapless spin-wave excitations because of the broken
continuous $O(3)$
symmetry of the ordered state. So we can no longer use the gap, $\Delta$, as
the energy
scale, $G=\Delta \sim (g_c - g)^{z\nu}$ for measuring deviations from $g=g_c$
for $g< g_c$. A
convenient substitute turns out to be the spin stiffness $G=\rho_s$ which has
the physical
dimensions of energy in $d=2$, and which also vanishes as $\rho_s \sim (g_c -
g)^{z\nu}$. At finite
$T$ in region A, it is known~\cite{bz,chn} that the spin correlation length
$\sim \exp(2 \pi \rho_s /T)$
as $T \rightarrow 0$. It is interesting to note that the behavior of ${\cal
H}_I$
in $d=1$ is very similar: in this case the correlation length is determined by
the
mean spacing between kinks, and therefore behaves as $\sim \exp(\Delta/T)$
for low $T$ in region A. (Related to the exponential divergence of the
correlation length, there
is a further sub-division of the ``ordered'' regime of region A
(Figs~\ref{isingfig}
and~\ref{isingomegafig}) at the energy scale $c /(\mbox{correlation length})$;
this complication
occurs for both the $d=2$ rotor and the
$d=1$ Ising models, and has been discussed elsewhere~\cite{csy,chn}.)

Associated with the continuous $O(3)$ symmetry of ${\cal H}_R$, there is an
important observable
whose properties cannot be deduced by an analogy with the the Ising model. This
is the uniform
susceptibility $\chi_H$, the response to a field, $H$ which couples to the
global conserved charge
associated with the continuous symmetry:
\begin{equation}
{\cal H}_R \rightarrow - H \sum_i L_{zi}
\end{equation}
The scaling dimension of $\chi_H$ can be determined exactly using symmetry
arguments and the
assumption of hyperscaling~\cite{cs,conserve}: this yields the scaling form
\begin{equation}
\chi_H = \frac{T}{c^2} \Phi_H \left (\frac{\rho_s}{T} \right)
\label{chiHscale}
\end{equation}
Here $c$, is the same velocity that appears in a $T=0$ correlator like
(\ref{chiteq0}),
and $\Phi_H$ is a fully universal function. This function was computed
exactly~\cite{cs,csy} in a
$O(N=\infty)$ rotor model, along with $1/N$ corrections in some limits; it has
the limiting behavior
\begin{equation}
\Phi_H(r ) = \left\{
\begin{array}{ll}
\displaystyle
\frac{\sqrt{5}}{\pi} \ln \left( \frac{\sqrt{5} + 1}{2} \right) \left[ 1 -
\frac{0.6189}{N}
+ \ldots \right] + \ldots & r \rightarrow 0 \\
\displaystyle \frac{2r}{N} + \frac{N-2}{N} + \ldots & r \rightarrow \infty
\end{array}\right.
\label{chiHres}
\end{equation}
The two terms in the second result ($r\rightarrow \infty$) are expected to be
exact to all orders
in $1/N$; the same two terms were also obtained by Hasenfratz and
Niedermayer~\cite{hasen}.
Notice from (\ref{chiHscale}) and (\ref{chiHres}) that $\chi_H$ has a linear
dependence on
$T$ both for $T \ll \rho_s$ and $T \gg \rho_s$; the slopes however differ by a
factor of about 3 (for
$N=3$) and this will be important for experimental comparisons~\cite{cs,csy}.

\subsection{Comparison with simulations and experiments on Heisenberg
antiferromagnets}
The most straightforward comparison is with the double-layer, spin-1/2
Heisenberg
antiferromagnet~\cite{double}. This model consists of spin-1/2 Heisenberg spins
on two adjacent square
lattices, with an intralayer antiferromagnetic exchange $J$ and an interlayer
antiferromagnetic exchange
$K$. The ratio
$K/J$ acts much like the dimensionless coupling $g$, with the large $K/J$ a
gapped quantum
paramagnet of singlet pairs of spins in opposite layers, and the small $K/J$
magnetically
ordered. Extensive numerical simulations have been carried out on this model by
Sandvik and
collaborators~\cite{sandvik}, and the critical point $K=K_c$ identified rather
precisely. It is then
possible to study the quantum-critical region C quite carefully as it extends
over the maximum $T$ range.
A number of universal amplitude ratios, including those associated with
$\chi_H$, and all
results are now in good agreement with the $1/N$ expansion on the $O(N)$
quantum rotor model.
Results from high temperature series expansions on the double-layer model also
support
this conclusion~\cite{elstner}.

Secondly, comparisons have been made with the single-layer, square lattice
spin-1/2 Heisenberg
antiferromagnet, both via simulations and by experimental measurements on $La_2
Cu O_4$.
This model has long-range order at $T=0$, so must map onto the non-linear sigma
model
with $g < g_c$. The low $T$ region A was studied in the paper of Chakravarty,
Halperin and
Nelson~\cite{chn}, with good experimental agreement. Here we focus on the issue
of whether this lattice
model exhibits the CQFT high $T$ behavior of region C, or it goes directly from
region A
to a non-universal, lattice dominated high $T$ region like D of
Fig~\ref{isingfig}.
It was first argued by Chubukov and Sachdev~\cite{cs} that this model does
indeed possess a significant
intermediate temperature regime of region C: this was based on comparisons with
the $T$ dependence
of $\chi_H$, with the factor of 3 alluded to above playing an important role.
They also
noted that it would be difficult to identify this region in the correlation
length,
an observation that was subsequently re-iterated by Greven {\em et.
al.\/}~\cite{greven}. A rather
convincing demonstration of the presence of region C was given recently by
Elstner {\em et.
al.}~\cite{elstner,rajiv} who examined a large number of observables in a high
$T$ expansion, and found
good consistency with the universal rotor model results.

Also significant in this context
have been nuclear magnetic resonance experiments of Imai {\em et.
al.}~\cite{imai} on $La_2 Cu O_4$.
They have
measured the $T$ dependence of the $1/T_1$ and $1/T_2$ relaxation rates at
intermediate
temperatures.
Their observations are in reasonable agreement with the
predictions~\cite{cs,csy,sokolpines,css} that can
be obtained from the universal scaling result for $\chi (k, \omega)$ in the
quantum-critical region C.

\section{Conclusions}
\label{concsec}
This paper has presented a discussion of the vicinity of a second-order quantum
phase transition in the
context of a number of simple models. The overall picture that emerges is
summarized in
Fig~\ref{generalfig}, which shows a generic phase diagram in the plane of a
coupling constant $g$, and
the temperature $T$ of a $d$-dimensional system; this phase diagram is a
generalization of
a diagram obtained first by Chakravarty {\em et. al.}~\cite{chn} for the $d=2$
quantum rotor model.
The quantum phase transition occurs at the point $g=g_c$, $T=0$. Associated
with this critical point,
we can define a continuum quantum field theory (CQFT) over spacetime.
In general, space ($x$) and time ($\tau$) do {\em not} play a similar role in
the CQFT and have different
scaling dimensions: $x \rightarrow x/s$, $\tau \rightarrow \tau/s^{z}$ under a
spatial rescaling
by $s$, with $z$ the dynamic exponent. Further, even in imaginary time, the
action for
the CQFT can be complex due to the presence of Berry phases, and therefore
corresponds to a statistical
mechanics model with complex weights.
Correlators of the CQFT are the
universal functions describing crossovers in the vicinity of $g=g_c$, $T=0$.
The CQFT has no ultraviolet
cutoff, but is characterized solely by two energy scales: the temperature $T$
and an energy
scale $G \sim |g - g_c|^{z\nu}$ characterizing the deviation of the ground
state from the
critical point (here $\nu$ is the correlation length exponent).
The value of the ratio $G/T$ determines two distinct regions of the CQFT shown
in Fig~\ref{generalfig}.
In both regions there is a high frequency regime ($\omega > \mbox{max} (T, G)$)
which is dominated
by excitations of the critical CQFT of the $g=g_c$, $T=0$ point. The regions
are distinguished
only by their low frequency behavior, which we discuss in turn:

\noindent ({\em i}) The low $T$ region
($T \ll G$, shown shaded in Fig~\ref{generalfig}) is a region of
``conventional'' physics for
frequency scales $\omega < G$.
This frequency regime can be understood by beginning with the non-critical
ground state and examining the
particle-like excitations above it. A simple classical model (a Boltzmann
equation for a
gas of quasiparticle excitations, or a Langevin model of the types
discussed by Hohenberg and Halperin~\cite{halphoh}) is usually adequate for
describing the
long-distance, long-time dynamics of these excitations. The shaded region can
also contain thermally
driven, phase transitions; these transitions will be described by a classical
field theory.

\noindent ({\em ii}) The high $T$ region of the CQFT has a novel quantum
relaxational regime.
This regime is not described by an effective classical model, and displays
intrinsic
quantum-mechanical effects at the longest time and distance scales.
A scaling analysis for this regime was reviewed in this paper. In some cases,
as in the model
of Section~\ref{isingsec}, it is possible to obtain an exact value for the
relaxation constant.

\begin{figure}
\epsfxsize=4.7in
\centerline{\epsffile{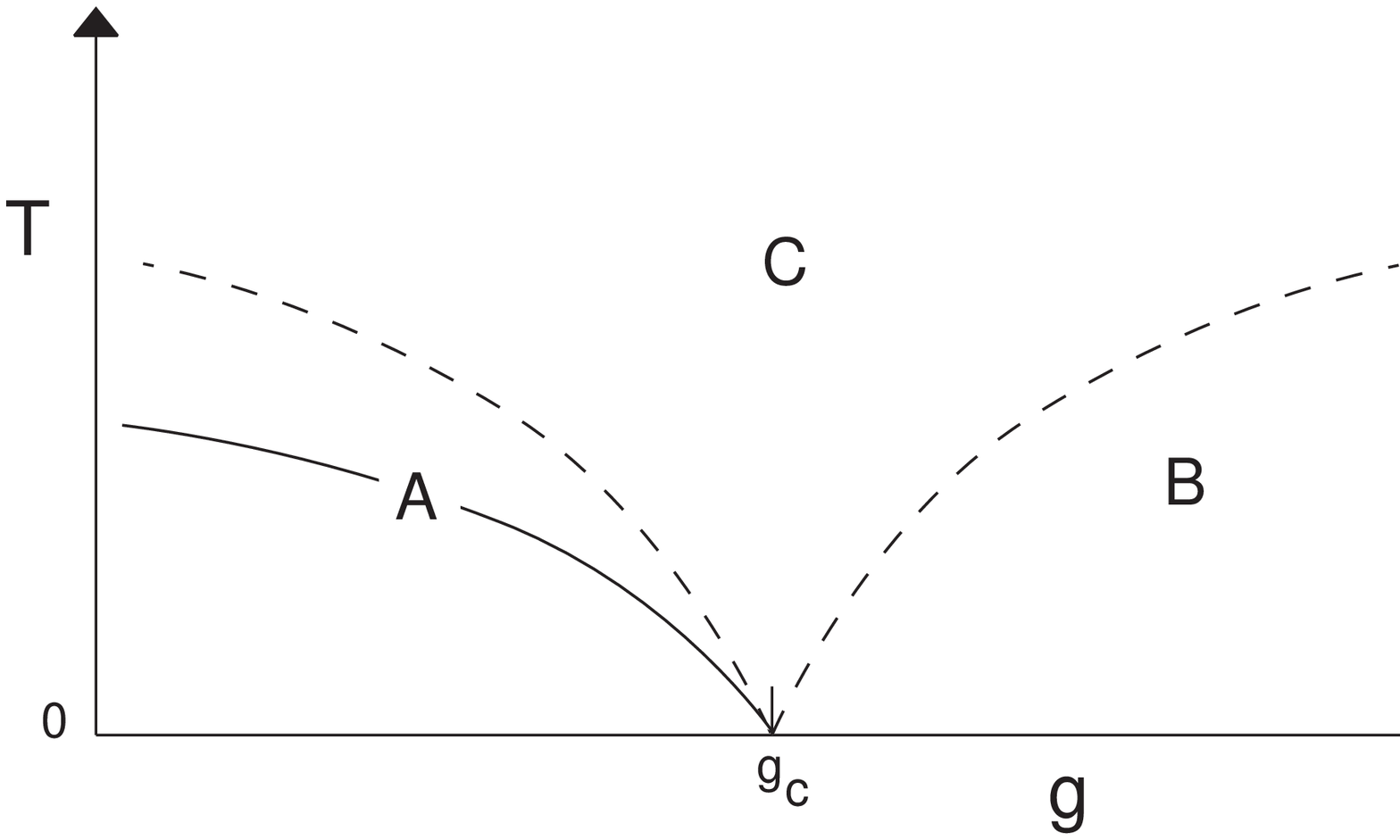}}
\vspace{0.05in}
\caption{Phase diagram of the transverse field Ising model in $d \geq 1$
dimensions; dashed lines
denote crossover boundaries, while the solid line is a phase transition. In
region A the
system displays characteristics of the ordered ground state; thermal
fluctuations about this
state can destroy long-range order at any non-zero $T$ in $d=1$, but for $d>1$
there is a
 phase transition (in the universality class of the $d$-dimensional classical
Ising model). In region B the system is a gapped quantum paramagnet, and region
C is
quantum-critical. The crossover boundaries obey $T \sim |g - g_c |^{z\nu}$ with
$z=1$
and $\nu$ the correlation length exponent of the $d+1$ dimensional classical
Ising model
($\nu=1$ in $d=1$).
The same phase diagram also applies to the quantum rotor model of
Section~\protect\ref{secrotor} in $d\geq 2$, with the difference that the phase
transition in region A
is present only for $d>2$ and is then in the universality class of the $d+1$
dimensional
classical Heisenberg ferromagnet. Chakravarty {\em et. al.}~\protect\cite{chn}
obtained this phase
diagram for the $d=2$ quantum rotor model.}
\label{isingfig}
%\end{figure}
%\begin{figure}
\vspace{0.2in}
\epsfxsize=4.7in
\centerline{\epsffile{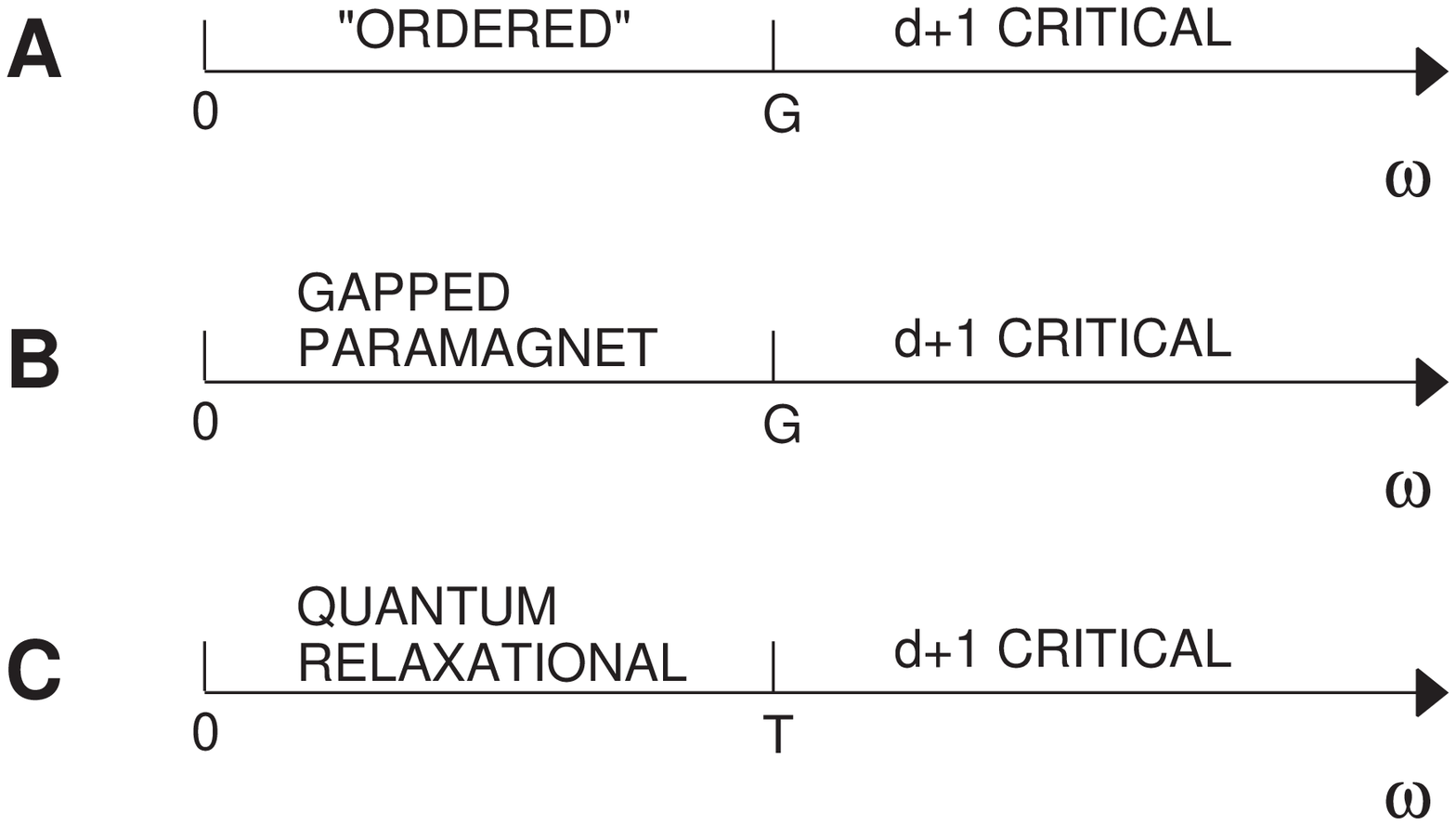}}
\vspace{0.05in}
\caption{Crossovers as a function of probe frequency in the regions of
Fig.~\protect\ref{isingfig} for
the transverse-field Ising model in $d$ dimensions.
The $d+1$ critical regime is described by the CQFT at $g=g_c$ and $T=0$. The
``ordered'' regime
is where a classical description in terms of a gas of well-separated domain
walls in an otherwise
ordered state is appropriate. In $d=1$ such ``kinks'' destroy long-range-order
at any non-zero $T$.
The same crossovers also apply to the quantum rotor model of
Section~\protect\ref{secrotor}: the
``ordered'' regime is now one where spin-wave fluctuations about an ordered
ground state dominate,
and these fluctuations destroy long-range-order at any non-zero $T$ for $d=2$.
}
\label{isingomegafig}
\end{figure}

\begin{figure}
\epsfxsize=5in
\centerline{\epsffile{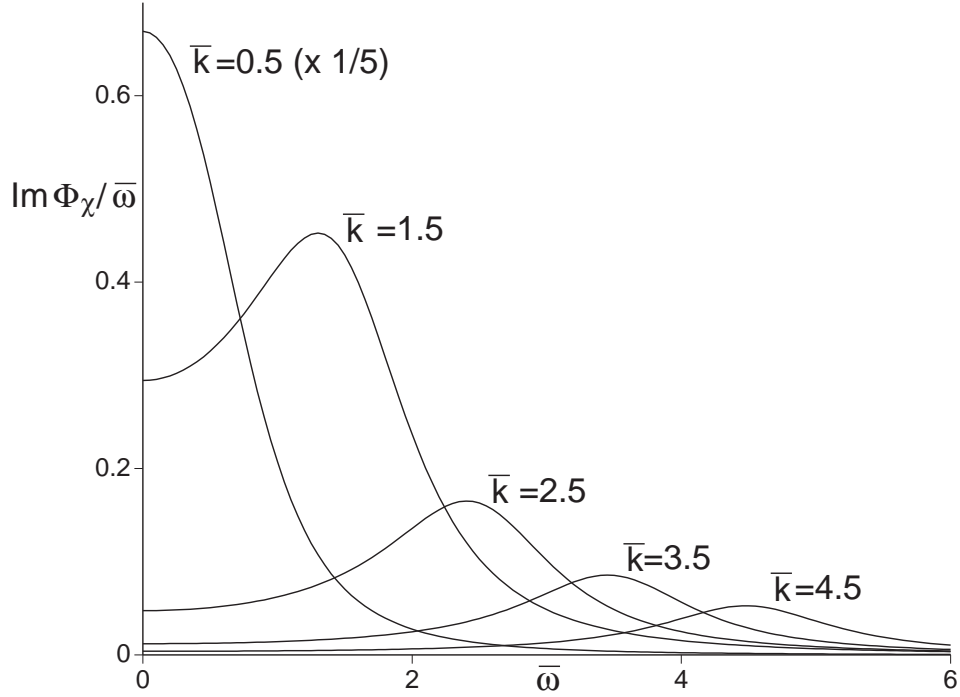}}
\vspace{0.1in}
\caption{Scaling function for the imaginary part of the dynamic susceptibility
in the quantum critical
region C of the transverse field Ising model in $d=1$. The susceptibility $\chi
= Z \Phi_\chi / T^{7/4}$
and $\kb=ck/T$, $\ob=\omega/T$.}
\label{chifig}
\end{figure}

\begin{figure}
\epsfxsize=4.5in
\centerline{\epsffile{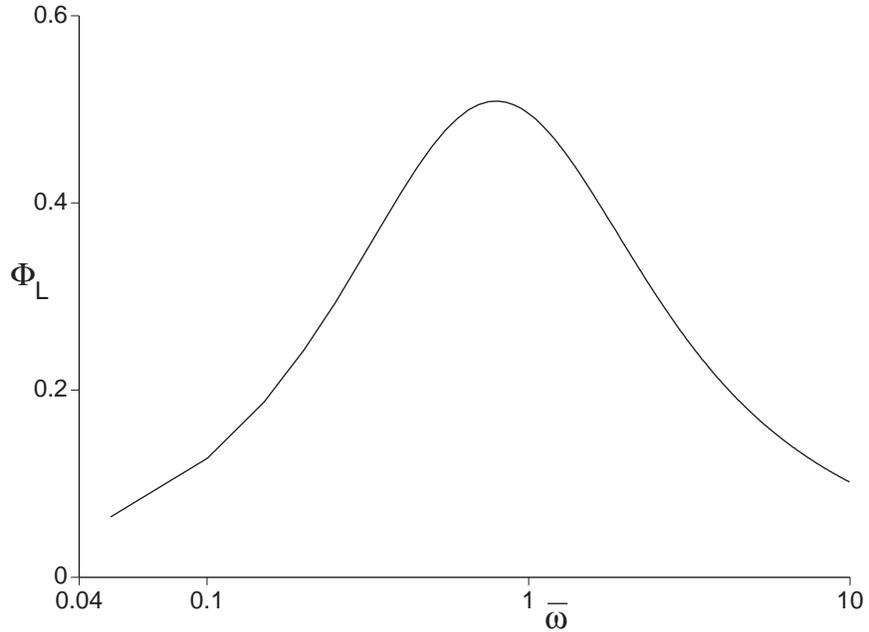}}
\vspace{0.1in}
\caption{Scaling function for the imaginary part of the local dynamic
susceptibility in the quantum
critical region C of the transverse field Ising model in $d=1$. The
susceptibility
$\chi_L^{\prime\prime}$ is related to $\Phi_L$ by (\protect\ref{phildef}),
$\ob=\omega/T$,
and notice the logarithmic scale on the horizontal axis.}
\label{chilocfig}
\end{figure}

\begin{figure}
\epsfxsize=4.5in
\centerline{\epsffile{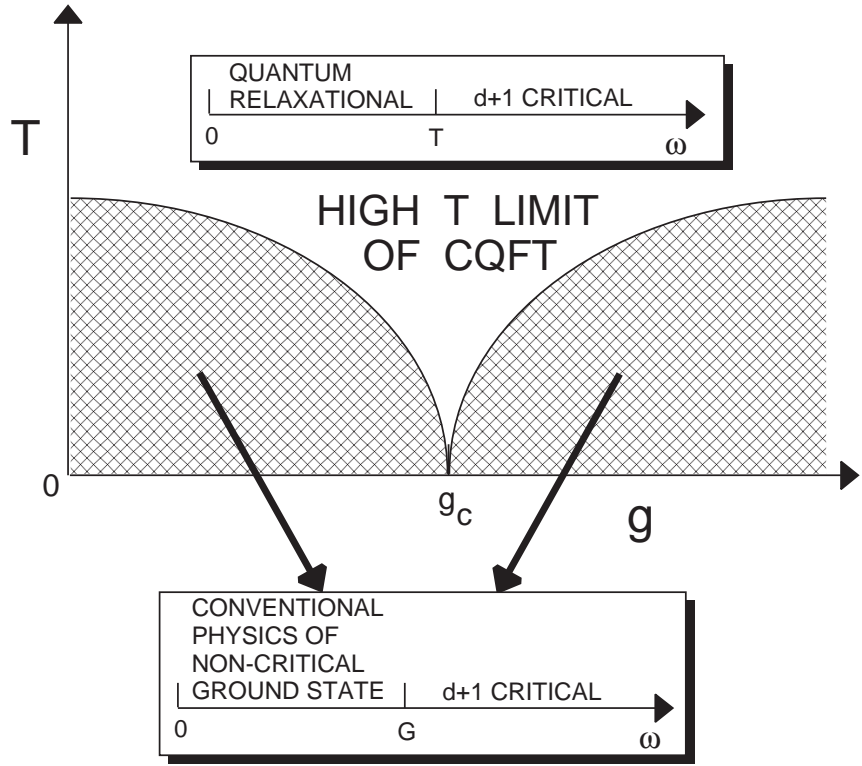}}
\vspace{0.1in}
\caption{Generic phase diagram of a $d$-dimensional system displaying a
second-order quantum transition
at $g=g_c$ and $T=0$. The boundary of the shaded region represents a crossover
and not a
phase transition. See the text
for more information.}
\label{generalfig}
\end{figure}

This paper has reviewed only a small portion of what is a rapidly developing
subject.
We list below a number of recent (and not so recent) developments in related
areas:
\begin{itemize}
\item Quantum transitions between Fermi liquids and states with various types
of spin or charge
density wave orderings were discussed in important early work by
Hertz~\cite{hertz}. He focussed
on the immediate vicinity of the finite termperature transition, like that
within region A
in Fig~\ref{isingfig}. In particular, Hertz missed the existence of the
``quantum-critical'' regime (as
was pointed out recently by Millis~\cite{millis}---this oversight is similar to
Suzuki's~\cite{suzuki} for the Ising model), which has been the main focus of
this paper.
More detailed studies of quantum transitions involving Fermi liquids have
appeared
recently~\cite{millis,ioffe,scs,georges,andrey,shankar2}. A related, but
different, perspective is
provided by studies of critical phenomena in rotor models with doped
electrons~\cite{ss,scs}.
\item Related ideas on scaling in the quantum critical region have been
presented by
Tsvelik and collaborators~\cite{tsvelik}.
\item A great deal of work has been done recently on ``quantum impurity''
models~\cite{affleck,varma}  like the multi-channel
Kondo effect. These models also display quantum phase transitions, with
crossovers bearing some
similarity to those discussed here. However
the transitions do not modify the bulk properties, and are related instead to
boundary critical
phenomena.
\item As we indicated briefly in the discussion on the Bose gas in
Section~\ref{sec:bose},
dangerously irrelevant operators sometimes need to be considered, as they do in
the dilute Bose gas for $d> 2$. In fact such effects arise somewhat more
frequently than
they do in classical critical phenomena, as the upper critical dimension of the
quantum
transition is often quite low. An early analysis of the dilute Bose gas in
$d=3$ by Weichmann {\em et. al.}~\cite{weichmann} was dominated by such
effects, although they
did not present their results in the general context of quantum phase
transitions.
Such a perspective can be found in more recent work~\cite{millis,oppermann}.
\item An important subject, on which much is not understood, is the effect of
quenched randomness
on quantum phase transitions.
An early analysis for the random quantum rotor model was given
by Boyanovsky and Cardy~\cite{boyanovsky}.
A recent exact solution by Fisher~\cite{daniel} of the random
transverse-field Ising model in $d=1$ represents significant progress. The
nature of spin glass
ordering at $T=0$ and its destruction by quantum fluctuations has also been
studied
recently~\cite{rsy,oppermann}.
\item A number of experiments~\cite{aeppli,keimer,aronson}
have reported scaling of the type in Eqn
(\ref{phildef}) in the local dynamic susceptibility. However the universality
classes controlling
these
systems are not understood and quenched randomness appears to play a
significant role.
Nevertheless, it is interesting to note the qualitative similarity between the
experimental
measurements in Fig.~4 of Aronson {\em et. al.}\cite{aronson} and our result
for the
local susceptibility in Fig~\ref{chilocfig}. The latter measures spin
correlations on a single
Ising spin induced via its coupling to its environment of other Ising spins.
However, there is a
fundamental equivalence between the spin being measured and its environment;
this is an
important difference between our bulk approach and alternative descriptions of
the
experiments using ``quantum impurity'' models~\cite{affleck,varma,aronson}
which clearly
distinguish between the impurity and environment degrees of freedom.

\item A recent study~\cite{ferro} has examined the CQFT of a quantum
ferromagnet. This system does not
display a quantum phase transition of the type discussed here. Nevertheless,
its phase diagram has
regions similar to those in Fig~\ref{generalfig} and aspects of its scaling
properties are related
to those of the dilute Bose gas of Section~\ref{sec:bose}.
\end{itemize}

\acknowledgments
I thank  N.~Read, R.~Shankar, A.~Sokol, T.~Senthil, J.~Ye, and especially
A.V.~Chubukov for
collaborations on the topics reviewed in this paper. I am grateful to
A.V.~Chubukov and
T.~Senthil for valuable comments on the manuscript and R.E. Shrock for helpful
discussions.
This research
was supported by National Science Foundation Grant DMR-9224290.

\end{document}